\documentclass[
]{ceurart}

\usepackage{listings}
\begin{document}

\copyrightyear{2023}
\copyrightclause{Copyright for this paper by its authors.
  Use permitted under Creative Commons License Attribution 4.0
  International (CC BY 4.0).}

\conference{HCMIR23: 2nd Workshop on Human-Centric Music Information Research, November 10th, 2023, Milan, Italy}

\title{TunesFormer: Forming Irish Tunes with Control Codes by Bar Patching}


\author{Shangda Wu}[%
email=shangda@mail.ccom.edu.cn,
]
\author{Xiaobing Li}[%
email=lxiaobing@ccom.edu.cn,
]
\author{Feng Yu}[%
email=yufeng@ccom.edu.cn,
]
\author{Maosong Sun}[%
email=sms@tsinghua.edu.cn,
]
\cormark[1]
\address{Department of Music AI and Information Technology, Central Conservatory of Music, Beijing, China}
\address{Department of Computer Science and Technology, Tsinghua University, Beijing, China}

\cortext[1]{Corresponding author.}


\begin{abstract}
This paper introduces TunesFormer, an efficient Transformer-based dual-decoder model specifically designed for the generation of melodies that adhere to user-defined musical forms. Trained on 214,122 Irish tunes, TunesFormer utilizes techniques including bar patching and control codes. Bar patching reduces sequence length and generation time, while control codes guide TunesFormer in producing melodies that conform to desired musical forms. Our evaluation demonstrates TunesFormer's superior efficiency, being 3.22 times faster than GPT-2 and 1.79 times faster than a model with linear complexity of equal scale while offering comparable performance in controllability and other metrics. TunesFormer provides a novel tool for musicians, composers, and music enthusiasts alike to explore the vast landscape of Irish music. Our model and code are available at \href{https://github.com/sander-wood/tunesformer}{GitHub}.
\end{abstract}


\maketitle

\section{Introduction}

Music composition is the art of crafting melodies and harmonies within structured forms, and recent neural networks have gained attention in generating melodies adhering to musical forms \cite{DBLP:journals/corr/abs-1811-08380,DBLP:conf/ijcnn/GuoMH21}. Yet, research mainly targets harmony, section lengths \cite{Naruse2022, DBLP:conf/mm/ZhangZQW022}, bar structure \cite{DBLP:journals/ai/WuLHZ20, https://doi.org/10.48550/arxiv.2110.05020}, or rules \cite{DBLP:conf/ismir/DaiJGD21, DBLP:journals/corr/abs-2208-14345}, missing a comprehensive understanding of musical form.

Melody generation is complex due to varied cultural traditions. Irish music, known for unique styles and instruments like the fiddle and tin whistle, uses forms and techniques like rolls and triplets for expression. Therefore, generating music like Irish folk tunes requires precise musical notation \cite{DBLP:journals/corr/SturmSBK16,geerlings2020interacting}. In addition, generating melodies with specific forms is challenging due to long-range dependencies. The Transformer model \cite{DBLP:conf/nips/VaswaniSPUJGKP17} addresses this by capturing these dependencies, but with a quadratic computational cost related to sequence length \cite{DBLP:conf/iclr/HuangVUSHSDHDE19,wu2023exploring}.

In response to the aforementioned challenges, we propose TunesFormer, a Transformer-based dual-decoder model that combines bar patching and control codes to efficiently generate expressive Irish music in ABC notation. Bar patching divides music into smaller segments, reducing sequence length and improving computational efficiency \cite{DBLP:journals/corr/abs-2304-11029}, while control codes enable users to define musical forms, allowing for controllable music generation.

The key contributions of this paper are as follows:

\begin{itemize}
    \item As a dual-decoder model based on bar patching, TunesFormer significantly accelerates generation speed while maintaining the quality of the generated music.
    \item TunesFormer enables users to generate melodies with diverse musical forms, providing flexibility and alignment with artistic vision through control codes.
    \item To support future research, we release the \textbf{Irish} \textbf{M}assive \textbf{A}BC \textbf{N}otation (IrishMAN) dataset, an open-source collection of 216,284 Irish tunes in the ABC notation format.
\end{itemize}

\section{Methodology}

\begin{figure*}[t]
	\centering
	\begin{minipage}{\textwidth} 
        \centering
        \includegraphics[width=\textwidth]{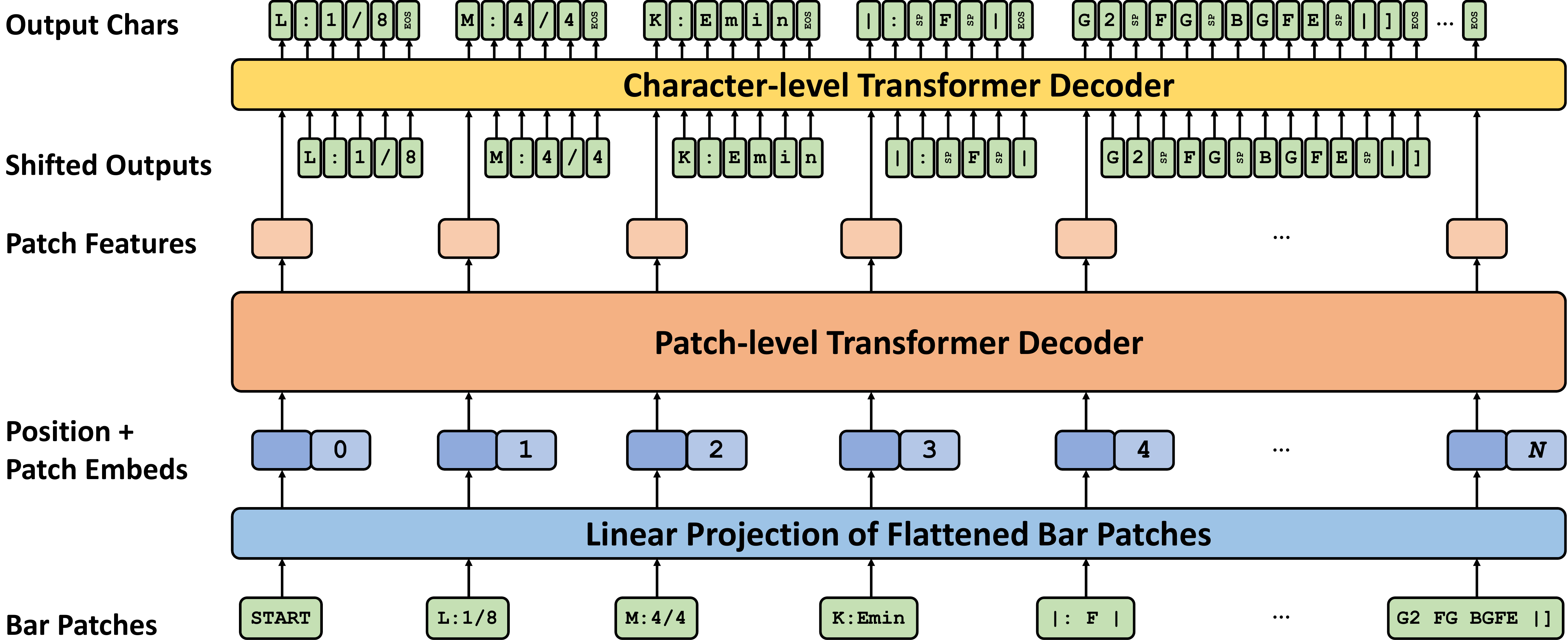}
	\end{minipage}
        \vspace{1em}
	\caption{The dual-decoder music generation framework is based on bar patching with a training objective of next patch prediction. The patch-level decoder captures the overall musical structures and themes of the melody, while the character-level decoder focuses on the details of each bar. For the purpose of demonstration, control codes and padding tokens are excluded.}
        \vspace{-1em}
 \end{figure*}

\subsection{TunesFormer}
TunesFormer uses bar patching \cite{DBLP:journals/corr/abs-2304-11029} for melody generation, leveraging the ABC notation format\footnote{\url{https://abcnotation.com/}} ideal for representing Irish music. Bar patching divides scores into segments, such as bars, shortening sequences and enhancing efficiency without sacrificing musical integrity.

Fig. 1 showcases TunesFormer's dual-decoder design. Bar patches are converted into embeddings that input to the patch-level decoder, producing patch features. These are input to the character-level decoder, which translates the patch features into the ABC notation sequences.

Given \(L\) as sequence length and \(P\) as patch size, bar patching reduces the patch-level decoder complexity from \(O(L^2)\) to \(O\left(\frac{L^2}{P^2}\right)\). Meanwhile, the character-level decoder complexity becomes \(O(LP)\). Considering \( M \) and \( N \) as parameter sizes for patch and character-level decoders respectively, computational need shifts from \( (M+N) \cdot L^2 \) to \( M \cdot \left(\frac{L^2}{P^2}\right) + N \cdot LP \). This is particularly advantageous for large sequences, high \( M \) to \( N \) ratios, and optimal \( P \) choices.

In our implementation, \(L=4096\), \(P=32\), yielding a 128 patch-length. The patch-level has 9 layers, and the character-level has 3, both with a 768 hidden size.

\subsection{Control Codes}
Inspired by CTRL \cite{DBLP:journals/corr/abs-1909-05858}, TunesFormer integrates control codes to denote musical forms. These codes precede the ABC notation, letting users dictate tune structures. Introduced codes are:

\begin{itemize}
\item \textbf{S:number of sections} - Dictates melody sections, ranging 1-8 (e.g., \texttt{S:1} for a single-section melody, and \texttt{S:8} for a melody with eight sections), based on symbols like \texttt{[|},\texttt{||},\texttt{|]},\texttt{|:},\texttt{::}, and \texttt{:|} used to represent section boundaries.
\item \textbf{B:number of bars} - Sets number of bars within a section. It counts on the bar symbol \texttt{|}. The range is 1 to 32 (e.g., \texttt{B:1} for a one-bar section, and \texttt{B:32} for a section with 32 bars).
\item \textbf{E:edit distance similarity} - Manages similarity between section $c$ and previous section $p$. Derived from Levenshtein distance \cite{levenshtein1966binary} $lev(c,p)$, it measures section differences:
\begin{equation}
eds(c,p) = 1 - \frac{lev(c,p)}{max(|c|,|p|)}
\end{equation}
where $|c|$ and $|p|$ are the string lengths of the two sections. It is discretized into 11 levels, ranging from 0 to 10 (e.g., \texttt{E:0} for no similarity, and \texttt{E:10} for an exact match). For the $N$-th section, there are $N-1$ previous sections to compare with.
\end{itemize}

While earlier methods leaned on hand-crafted rules or limited training data \cite{DBLP:journals/corr/abs-2208-14345, DBLP:conf/ismir/JiangCZX22}, our control codes directly extract precise musical form information from ABC notation, thus leveraging large datasets to improve understanding of musical structures.

\subsection{Dataset}
The \textbf{IrishMAN} dataset\footnote{\url{https://huggingface.co/datasets/sander-wood/irishman}} has 216,284 Irish ABC tunes. 99\% (214,122) are for training and 1\% (2,162) for validation, sourced from thesession.org and abcnotation.com. Uniformity is maintained by converting tunes to XML and back using scripts\footnote{\url{https://wim.vree.org/svgParse/}}, with natural language fields removed.

Tunes have control codes from ABC symbols (Section 2.2) indicating musical forms. The music21-filtered subset\cite{DBLP:conf/ismir/CuthbertA10} contains 34,211 human-annotated lead sheets. This subset helped TunesFormer generate harmonized melodies. In addition, all tunes are public domain, ensuring ethical and legal use for research and creative projects.

\section{Experiments}
In the experiments, we used baselines like LSTM \cite{DBLP:journals/corr/SturmSBK16} for generating ABC notation, GPT-2 \cite{radford2019language} for music generation \cite{DBLP:conf/iclr/HuangVUSHSDHDE19,wu2023exploring}, and RWKV \cite{DBLP:journals/corr/abs-2305-13048}, which rivals Transformers in performance. All models were trained on the same IrishMAN dataset split with character-level ABC tokenization, using random sampling for decoding. The evaluation involved two objective metrics based on 1,000 tunes generated from scratch per model:

\begin{table*}[t!]
    \textbf{Table 1} \quad Comparison of different models in terms of performance across various metrics.
    \centering
    \resizebox{\linewidth}{!}{\begin{tabular}{p{2.1cm}p{2.1cm}<{\centering}p{1.6cm}<{\centering}p{2.3cm}<{\centering}p{2.4cm}<{\centering}p{1.8cm}<{\centering}p{2.4cm}<{\centering}p{2cm}<{\centering}}
      \toprule 
      \textit{Model} & Parameters & Efficiency & Controllability & Engagement & Authenticity & Harmoniousness & Playability \\
      \midrule 
      \textit{LSTM \cite{DBLP:journals/corr/SturmSBK16}} & 5,582,976 & 186.42 & 0.8931 & 0.1923 & 0.3462 & 0.2692 & 0.3462\\
      \textit{GPT-2 \cite{radford2019language}} & 85,940,736 & 84.32 & 0.9479 & 0.4231 & \textbf{0.4615} & \textbf{0.4615} & \textbf{0.4231}\\
      \textit{RWKV \cite{DBLP:journals/corr/abs-2305-13048}} & 82,057,728 & 151.46 & \textbf{0.9634} & \textbf{0.4615} & 0.4231 & \textbf{0.4615} & 0.3846\\
      \textit{TunesFormer} & 88,425,984 & \textbf{271.13} & 0.9584 & 0.3462 & \textbf{0.4615} & 0.3846 & \textbf{0.4231}\\
      \bottomrule 
    \end{tabular}}
    \vspace{-1em}
\end{table*}

\begin{itemize}
\item \textbf{Efficiency:} The number of tokens generated per second on an RTX 2080 Ti.
\item \textbf{Controllability:} Quantifying control precision by comparing edit distance between generated and actual control codes.
\end{itemize}

We used comparative evaluations due to the inconsistency in human values. Thirteen Irish musicians compared melody pairs: one from thesession.org with chord symbols, and a model-generated continuation from the initial two bars. Tune choice and order were randomized to avoid bias. Participants selected the melody that best aligned with the below descriptions:

\begin{itemize}
    \item \textbf{Engagement:} Captivating to the ear, evokes emotional resonance, and maintains the listener's interest.
    \item \textbf{Authenticity:} Representing the distinctive characteristics of Irish traditional music.
    \item \textbf{Harmoniousness:} Creating a natural flow that unifies melody and harmony into a cohesive and pleasing musical experience.
    \item \textbf{Playability:} Well-suited for performance and offers a wide range of playing techniques.
\end{itemize}

Participants chose between three options for each melody pair: 0 for human-composed, 1 for model-generated, and 0.5 for no preference. Thus, scores ranged from 0 to 1. Participants were instructed to skip melodies they were already familiar with to avoid bias.

Table 1 shows the evaluation of music generation models. TunesFormer, with 88,425,984 parameters and a Transformer base, is 3.22 times faster than GPT-2 and 1.79 times faster than RWKV. Its dual-decoder architecture focuses on character generation, explaining its efficiency despite its large size. It is worth highlighting that TunesFormer's efficiency does not come at the expense of its performance. Particularly noteworthy is its remarkable controllability, matching the highest scores achieved in authenticity and playability. The performance is enhanced by the interaction between the patch-level and character-level decoders, where the former contextualizes bar features, enabling the latter to create coherent compositions. In essence, TunesFormer's dual-decoder design boosts efficiency in melody generation without sacrificing quality, and shows a significant advantage over its competitors in the field.

\section{Conclusions}
This paper presents TunesFormer, a model that generates melodies using control codes and bar patching. The use of control codes enhances user interaction, enabling personalized and customizable music generation. The dual-decoder architecture employed by TunesFormer, combined with its bar patching mechanism, yields significant improvements in generation speed without compromising the quality of the generated music. Future directions include incorporating more musical features and applying TunesFormer to various cultural traditions.

\section*{Acknowledgments}
The authors gratefully acknowledge the financial support from the Special Program of National Natural Science Foundation of China (Grant No. T2341003), the Advanced Discipline Construction Project of Beijing Universities, the Major Program of National Social Science Fund of China (Grant No. 21ZD19), and the Nation Culture and Tourism Technological Innovation Engineering Project (Research and Application of 3D Music).

\bibliography{sample-ceur}

\begin{thebibliography}{20}
\expandafter\ifx\csname natexlab\endcsname\relax\def\natexlab#1{#1}\fi
\providecommand{\url}[1]{\texttt{#1}}
\providecommand{\href}[2]{#2}
\providecommand{\path}[1]{#1}
\providecommand{\DOIprefix}{doi:}
\providecommand{\ArXivprefix}{arXiv:}
\providecommand{\URLprefix}{URL: }
\providecommand{\Pubmedprefix}{pmid:}
\providecommand{\doi}[1]{\href{http://dx.doi.org/#1}{\path{#1}}}
\providecommand{\Pubmed}[1]{\href{pmid:#1}{\path{#1}}}
\providecommand{\bibinfo}[2]{#2}
\ifx\xfnm\relax \def\xfnm[#1]{\unskip,\space#1}\fi
\bibitem[{Chen et~al.(2018)Chen, Zhang, Dubnov, and
  Xia}]{DBLP:journals/corr/abs-1811-08380}
\bibinfo{author}{K.~Chen}, \bibinfo{author}{W.~Zhang},
  \bibinfo{author}{S.~Dubnov}, \bibinfo{author}{G.~Xia},
\newblock \bibinfo{title}{The effect of explicit structure encoding of deep
  neural networks for symbolic music generation},
\newblock \bibinfo{journal}{CoRR}  (\bibinfo{year}{2018}).
  \href{http://arxiv.org/abs/1811.08380}{{\tt arXiv:1811.08380}}.
\bibitem[{Guo et~al.(2021)Guo, Makris, and Herremans}]{DBLP:conf/ijcnn/GuoMH21}
\bibinfo{author}{Z.~Guo}, \bibinfo{author}{D.~Makris},
  \bibinfo{author}{D.~Herremans},
\newblock \bibinfo{title}{Hierarchical recurrent neural networks for
  conditional melody generation with long-term structure},
\newblock in: \bibinfo{booktitle}{International Joint Conference on Neural
  Networks, {IJCNN} 2021, Shenzhen, China, July 18-22, 2021},
  \bibinfo{publisher}{{IEEE}}, \bibinfo{year}{2021}.
  \DOIprefix\doi{10.1109/IJCNN52387.2021.9533493}.
\bibitem[{Naruse et~al.(2022)Naruse, Takahata, Mukuta, and Harada}]{Naruse2022}
\bibinfo{author}{D.~Naruse}, \bibinfo{author}{T.~Takahata},
  \bibinfo{author}{Y.~Mukuta}, \bibinfo{author}{T.~Harada},
\newblock \bibinfo{title}{Pop music generation with controllable phrase
  lengths},
\newblock in: \bibinfo{booktitle}{Proc. of the 23rd Int. Society for Music
  Information Retrieval Conf.}, \bibinfo{address}{Bengaluru, India},
  \bibinfo{year}{2022}.
\bibitem[{Zhang et~al.(2022)Zhang, Zhang, Qiu, Wang, and
  Zhou}]{DBLP:conf/mm/ZhangZQW022}
\bibinfo{author}{X.~Zhang}, \bibinfo{author}{J.~Zhang},
  \bibinfo{author}{Y.~Qiu}, \bibinfo{author}{L.~Wang},
  \bibinfo{author}{J.~Zhou},
\newblock \bibinfo{title}{Structure-enhanced pop music generation via
  harmony-aware learning},
\newblock in: \bibinfo{booktitle}{{MM} '22: The 30th {ACM} International
  Conference on Multimedia, Lisboa, Portugal, October 10 - 14, 2022},
  \bibinfo{publisher}{{ACM}}, \bibinfo{year}{2022}.
  \DOIprefix\doi{10.1145/3503161.3548084}.
\bibitem[{Wu et~al.(2020)Wu, Liu, Hu, and Zhu}]{DBLP:journals/ai/WuLHZ20}
\bibinfo{author}{J.~Wu}, \bibinfo{author}{X.~Liu}, \bibinfo{author}{X.~Hu},
  \bibinfo{author}{J.~Zhu},
\newblock \bibinfo{title}{Popmnet: Generating structured pop music melodies
  using neural networks},
\newblock \bibinfo{journal}{Artif. Intell.}  (\bibinfo{year}{2020}).
  \DOIprefix\doi{10.1016/j.artint.2020.103303}.
\bibitem[{Zou et~al.(2021)Zou, Zou, Zhao, Zhang, Zhang, and
  Wang}]{https://doi.org/10.48550/arxiv.2110.05020}
\bibinfo{author}{Y.~Zou}, \bibinfo{author}{P.~Zou}, \bibinfo{author}{Y.~Zhao},
  \bibinfo{author}{K.~Zhang}, \bibinfo{author}{R.~Zhang},
  \bibinfo{author}{X.~Wang}, \bibinfo{title}{Melons: generating melody with
  long-term structure using transformers and structure graph},
  \bibinfo{year}{2021}. \URLprefix \url{https://arxiv.org/abs/2110.05020}.
  \DOIprefix\doi{10.48550/ARXIV.2110.05020}.
\bibitem[{Dai et~al.(2021)Dai, Jin, Gomes, and
  Dannenberg}]{DBLP:conf/ismir/DaiJGD21}
\bibinfo{author}{S.~Dai}, \bibinfo{author}{Z.~Jin}, \bibinfo{author}{C.~Gomes},
  \bibinfo{author}{R.~B. Dannenberg},
\newblock \bibinfo{title}{Controllable deep melody generation via hierarchical
  music structure representation},
\newblock in: \bibinfo{booktitle}{Proceedings of the 22nd International Society
  for Music Information Retrieval Conference, {ISMIR} 2021, Online, November
  7-12, 2021}, \bibinfo{year}{2021}.
\bibitem[{Lu et~al.(2022)Lu, Tan, Yu, Qin, Zhao, and
  Liu}]{DBLP:journals/corr/abs-2208-14345}
\bibinfo{author}{P.~Lu}, \bibinfo{author}{X.~Tan}, \bibinfo{author}{B.~Yu},
  \bibinfo{author}{T.~Qin}, \bibinfo{author}{S.~Zhao},
  \bibinfo{author}{T.~Liu},
\newblock \bibinfo{title}{Meloform: Generating melody with musical form based
  on expert systems and neural networks},
\newblock \bibinfo{journal}{CoRR}  (\bibinfo{year}{2022}).
  \href{http://arxiv.org/abs/2208.14345}{{\tt arXiv:2208.14345}}.
\bibitem[{Sturm et~al.(2016)Sturm, Santos, Ben{-}Tal, and
  Korshunova}]{DBLP:journals/corr/SturmSBK16}
\bibinfo{author}{B.~L. Sturm}, \bibinfo{author}{J.~F. Santos},
  \bibinfo{author}{O.~Ben{-}Tal}, \bibinfo{author}{I.~Korshunova},
\newblock \bibinfo{title}{Music transcription modelling and composition using
  deep learning},
\newblock \bibinfo{journal}{CoRR}  (\bibinfo{year}{2016}).
  \href{http://arxiv.org/abs/1604.08723}{{\tt arXiv:1604.08723}}.
\bibitem[{Geerlings and Merono-Penuela(2020)}]{geerlings2020interacting}
\bibinfo{author}{C.~Geerlings}, \bibinfo{author}{A.~Merono-Penuela},
\newblock \bibinfo{title}{Interacting with gpt-2 to generate controlled and
  believable musical sequences in abc notation},
\newblock in: \bibinfo{booktitle}{Proceedings of the 1st Workshop on NLP for
  Music and Audio (NLP4MusA)}, \bibinfo{year}{2020}.
\bibitem[{Vaswani et~al.(2017)Vaswani, Shazeer, Parmar, Uszkoreit, Jones,
  Gomez, Kaiser, and Polosukhin}]{DBLP:conf/nips/VaswaniSPUJGKP17}
\bibinfo{author}{A.~Vaswani}, \bibinfo{author}{N.~Shazeer},
  \bibinfo{author}{N.~Parmar}, \bibinfo{author}{J.~Uszkoreit},
  \bibinfo{author}{L.~Jones}, \bibinfo{author}{A.~N. Gomez},
  \bibinfo{author}{L.~Kaiser}, \bibinfo{author}{I.~Polosukhin},
\newblock \bibinfo{title}{Attention is all you need},
\newblock in: \bibinfo{booktitle}{Advances in Neural Information Processing
  Systems 30: Annual Conference on Neural Information Processing Systems 2017,
  December 4-9, 2017, Long Beach, CA, {USA}}, \bibinfo{year}{2017}.
\bibitem[{Huang et~al.(2019)Huang, Vaswani, Uszkoreit, Simon, Hawthorne,
  Shazeer, Dai, Hoffman, Dinculescu, and Eck}]{DBLP:conf/iclr/HuangVUSHSDHDE19}
\bibinfo{author}{C.~A. Huang}, \bibinfo{author}{A.~Vaswani},
  \bibinfo{author}{J.~Uszkoreit}, \bibinfo{author}{I.~Simon},
  \bibinfo{author}{C.~Hawthorne}, \bibinfo{author}{N.~Shazeer},
  \bibinfo{author}{A.~M. Dai}, \bibinfo{author}{M.~D. Hoffman},
  \bibinfo{author}{M.~Dinculescu}, \bibinfo{author}{D.~Eck},
\newblock \bibinfo{title}{Music transformer: Generating music with long-term
  structure},
\newblock in: \bibinfo{booktitle}{7th International Conference on Learning
  Representations, {ICLR} 2019, New Orleans, LA, USA, May 6-9, 2019},
  \bibinfo{publisher}{OpenReview.net}, \bibinfo{year}{2019}.
\bibitem[{Wu and Sun(2023)}]{wu2023exploring}
\bibinfo{author}{S.~Wu}, \bibinfo{author}{M.~Sun},
\newblock \bibinfo{title}{Exploring the efficacy of pre-trained checkpoints in
  text-to-music generation task},
\newblock in: \bibinfo{booktitle}{The AAAI-23 Workshop on Creative AI Across
  Modalities}, \bibinfo{year}{2023}. \URLprefix
  \url{https://openreview.net/forum?id=QmWXskBhesn}.
\bibitem[{Wu et~al.(2023)Wu, Yu, Tan, and
  Sun}]{DBLP:journals/corr/abs-2304-11029}
\bibinfo{author}{S.~Wu}, \bibinfo{author}{D.~Yu}, \bibinfo{author}{X.~Tan},
  \bibinfo{author}{M.~Sun},
\newblock \bibinfo{title}{Clamp: Contrastive language-music pre-training for
  cross-modal symbolic music information retrieval},
\newblock \bibinfo{journal}{CoRR}  (\bibinfo{year}{2023}). \URLprefix
  \url{https://doi.org/10.48550/arXiv.2304.11029}.
  \DOIprefix\doi{10.48550/arXiv.2304.11029}.
  \href{http://arxiv.org/abs/2304.11029}{{\tt arXiv:2304.11029}}.
\bibitem[{Keskar et~al.(2019)Keskar, McCann, Varshney, Xiong, and
  Socher}]{DBLP:journals/corr/abs-1909-05858}
\bibinfo{author}{N.~S. Keskar}, \bibinfo{author}{B.~McCann},
  \bibinfo{author}{L.~R. Varshney}, \bibinfo{author}{C.~Xiong},
  \bibinfo{author}{R.~Socher},
\newblock \bibinfo{title}{{CTRL:} {A} conditional transformer language model
  for controllable generation},
\newblock \bibinfo{journal}{CoRR}  (\bibinfo{year}{2019}). \URLprefix
  \url{http://arxiv.org/abs/1909.05858}.
  \href{http://arxiv.org/abs/1909.05858}{{\tt arXiv:1909.05858}}.
\bibitem[{Levenshtein et~al.(1966)}]{levenshtein1966binary}
\bibinfo{author}{V.~I. Levenshtein}, et~al.,
\newblock \bibinfo{title}{Binary codes capable of correcting deletions,
  insertions, and reversals},
\newblock in: \bibinfo{booktitle}{Soviet physics doklady},
  \bibinfo{organization}{Soviet Union}, \bibinfo{year}{1966}.
\bibitem[{Jiang et~al.(2022)Jiang, Chin, Zhang, and
  Xia}]{DBLP:conf/ismir/JiangCZX22}
\bibinfo{author}{J.~Jiang}, \bibinfo{author}{D.~Chin},
  \bibinfo{author}{Y.~Zhang}, \bibinfo{author}{G.~Xia},
\newblock \bibinfo{title}{Learning hierarchical metrical structure beyond
  measures},
\newblock in: \bibinfo{booktitle}{Proceedings of the 23rd International Society
  for Music Information Retrieval Conference, {ISMIR} 2022, Bengaluru, India,
  December 4-8, 2022}, \bibinfo{year}{2022}. \URLprefix
  \url{https://archives.ismir.net/ismir2022/paper/000023.pdf}.
\bibitem[{Cuthbert and Ariza(2010)}]{DBLP:conf/ismir/CuthbertA10}
\bibinfo{author}{M.~S. Cuthbert}, \bibinfo{author}{C.~Ariza},
\newblock \bibinfo{title}{Music21: {A} toolkit for computer-aided musicology
  and symbolic music data},
\newblock \bibinfo{publisher}{International Society for Music Information
  Retrieval}, \bibinfo{year}{2010}.
\bibitem[{Radford et~al.(2019)Radford, Wu, Child, Luan, Amodei, Sutskever
  et~al.}]{radford2019language}
\bibinfo{author}{A.~Radford}, \bibinfo{author}{J.~Wu},
  \bibinfo{author}{R.~Child}, \bibinfo{author}{D.~Luan},
  \bibinfo{author}{D.~Amodei}, \bibinfo{author}{I.~Sutskever}, et~al.,
\newblock \bibinfo{title}{Language models are unsupervised multitask learners},
\newblock \bibinfo{journal}{OpenAI blog}  (\bibinfo{year}{2019}).
\bibitem[{Peng et~al.(2023)Peng, Alcaide, Anthony, Albalak, Arcadinho, Cao,
  Cheng, Chung, Grella, V., He, Hou, Kazienko, Kocon, Kong, Koptyra, Lau,
  Mantri, Mom, Saito, Tang, Wang, Wind, Wozniak, Zhang, Zhang, Zhao, Zhou, Zhu,
  and Zhu}]{DBLP:journals/corr/abs-2305-13048}
\bibinfo{author}{B.~Peng}, \bibinfo{author}{E.~Alcaide},
  \bibinfo{author}{Q.~Anthony}, \bibinfo{author}{A.~Albalak},
  \bibinfo{author}{S.~Arcadinho}, \bibinfo{author}{H.~Cao},
  \bibinfo{author}{X.~Cheng}, \bibinfo{author}{M.~Chung},
  \bibinfo{author}{M.~Grella}, \bibinfo{author}{K.~K.~G. V.},
  \bibinfo{author}{X.~He}, \bibinfo{author}{H.~Hou},
  \bibinfo{author}{P.~Kazienko}, \bibinfo{author}{J.~Kocon},
  \bibinfo{author}{J.~Kong}, \bibinfo{author}{B.~Koptyra},
  \bibinfo{author}{H.~Lau}, \bibinfo{author}{K.~S.~I. Mantri},
  \bibinfo{author}{F.~Mom}, \bibinfo{author}{A.~Saito},
  \bibinfo{author}{X.~Tang}, \bibinfo{author}{B.~Wang}, \bibinfo{author}{J.~S.
  Wind}, \bibinfo{author}{S.~Wozniak}, \bibinfo{author}{R.~Zhang},
  \bibinfo{author}{Z.~Zhang}, \bibinfo{author}{Q.~Zhao},
  \bibinfo{author}{P.~Zhou}, \bibinfo{author}{J.~Zhu},
  \bibinfo{author}{R.~Zhu},
\newblock \bibinfo{title}{{RWKV:} reinventing rnns for the transformer era},
\newblock \bibinfo{journal}{CoRR}  (\bibinfo{year}{2023}). \URLprefix
  \url{https://doi.org/10.48550/arXiv.2305.13048}.
  \DOIprefix\doi{10.48550/arXiv.2305.13048}.
  \href{http://arxiv.org/abs/2305.13048}{{\tt arXiv:2305.13048}}.

\end{thebibliography}

\end{document}